\documentclass[twocolumn,showpacs,preprintnumbers,amsmath,amssymb]{revtex4}
\usepackage{graphicx}
\usepackage{amsmath}
 \usepackage{epstopdf}
\usepackage{color}

\begin{document}
\title{The excitation of rogue waves in a variable medium: an experimental study 
on the interaction of water waves and currents}

\author{A. Toffoli$^{1,2}$}
\author{T. Waseda$^3$}
\author{H. Houtani$^{3,4}$}
\author{T. Kinoshita$^{5}$}
\author{K. Collins$^2$}
\author{D. Proment$^6$}
\author{M. Onorato$^{7,8}$}

\affiliation{
$^1$Centre for Ocean Engineering Science and Technology, Swinburne 
University of Technology, P.O. Box 218, Hawthorn, 3122 Vic., Australia;\\
$^2$School of Marine Science and Engineering, Plymouth University, Plymouth
 PL4 8AA, UK;\\
$^3$Graduate School of Frontier Sciences, University
of Tokyo, Kashiwa, Chiba 277-8563, Japan;\\
$^4$National Maritime Research Institute, Shinkawa, Mitaka-shi, Tokyo 180-0004, Japan;\\
$^5$Institute of Industrial Science, University of Tokyo, Tokyo, Japan;\\
$^6$School of Mathematics, University of East Anglia, Norwich, NR4 7TJ, UK;\\
$^7$Dipartimento di Fisica, Universit{\`a} degli Studi di Torino, Via Pietro Giuria 1, 10125 Torino, Italy\\
$^8$INFN, Sezione di Torino, Via Pietro Giuria 1, 10125 Torino, Italy.
}

\date{\today}

\begin{abstract}
We show experimentally that a stable wave propagating into a region characterized by an opposite current may become modulationaly unstable.
 Experiments have been performed in two independent wave tank facilities; both of them are equipped with a wavemaker 
 and a  pump for generating a current propagating in the  opposite direction 
 with respect to the waves.
 The experimental results support a recent conjecture based on a current-modified Nonlinear 
 Schr\"odinger equation which establishes  
 that rogue waves can be triggered by non-homogeneous  
  current characterized by a negative horizontal velocity gradient. 
\end{abstract}
\maketitle

Ocean waves are characterized by a statistically small steepness and often
(but not always, see for example 
\cite{zakharov2010shape}) a weakly nonlinear approach is sufficient to capture some of the intriguing aspects hidden in the  fully nonlinear primitive equations. This weakly nonlinear approach is also shared by other fields of physics 
such as for example nonlinear optics \cite{newell1992nonlinear} and plasma physics \cite{zakharov1985hamiltonian} 
where small parameters can be individuated and asymptotic expansions can be used to simplify the original equations. If the considered physical process is not only weakly nonlinear but  also narrow-banded then the  lion's share is played by the Nonlinear Schr\"odinger  equation (NLS). Being an exactly integrable equation {\it via} the Inverse Scattering Transform
\cite{zakharov1974scheme}, bizarre analytical solutions have been found in the past: besides traveling waves,
breathers or multi-breather solutions have been found \cite{kuznetsov77,akhmediev87,peregrine1983water}
and observed in hydrodynamic  \cite{chabchoub2011rogue,chabchoub2012super}, nonlinear optics \cite{kibler2012observation,kibler2010peregrine} and plasma  \cite{Bailung2011observation} experiments. Starting from \cite{dysthe99,osborne00},
such solutions have been considered as prototypes of rogue waves. 
The early stages of 
the so called {\it Akhmediev breather} solution \cite{akhmediev87} describes the 
exponential growth of a slightly perturbed plane waves, i.e., it corresponds to the 
classical modulational instability process \cite{zakharov2009}. For water waves in infinite water depth, 
the instability is active when $\varepsilon N \geq 1 / \sqrt{2}$, where $\varepsilon = k_0 a_0$ is the initial steepness of the plane wave, $k_0$ the wavenumber, $a_0$ its amplitude and $N=\omega_0 / \Delta \Omega$ is the number of waves under the modulation; $\omega_0$ is the angular frequency corresponding to the carrier wavenumber $k_0$ and $\Delta \Omega$ the angular frequency of the modulation.

The whole picture is by now pretty well understood and relies on the fact that the medium in which waves
propagate is homogeneous. In terms of the NLS equation this means that the coefficients of 
the dispersive and nonlinear terms do not depend on the spatial coordinates. Much more complicated 
and intriguing  is the case in which the  medium changes its properties along the direction 
of propagation of the waves. This situation is much more difficult to treat in terms of simplified models
because it turns out that in general the resulting modified  NLS does not share the
 property of integrability as the standard NLS and analytical breathers  solutions can be found only in special cases (see some examples in  \cite{yan2010nonautonomous,dai2011nonautonomous,onorato2012approximate}). 

In the oceanographic context, the non-homogeneity of the medium is mostly due to 
currents or bottom topography. In this Letter we will consider, from an
experimental point of view, the interaction of 
waves and currents and the consequent formation of rogue waves.
Wave-current interaction and its effect on wave instability have been the subject of a number studies over the past decades \cite{chawla00,chawla02,gerber87,lai89,moreira12,stocker99,suastika04,toffoli11}. Only recently, however, has the rogue wave formation process induced by an opposite current gradient been studied numerically on 
the basis of the modified NLS equation \citep{hjelmervik09,onorato2011,ying2011linear,ruban2012nonlinear}. Specifically, \cite{onorato2011}  applied a transformation to the one dimensional current-modified NLS equation derived  in \cite{hjelmervik09} to obtain the following NLS-type of equation:
\begin{equation}
\frac{\partial B}{\partial x}+i \frac{ k_0}{\omega_0^2}\frac{\partial^2 B}{\partial t^2}
+i k_0^3\exp\left(-\Delta U/c_{g}\right)|B|^2B =0, \label{NLSM}
\end{equation}
where $c_{g}$ is the group velocity, $\Delta U=U(x)-U(0)$, with $U(x)$ the velocity of the current at position $x$ and U(0) is the current at $x=0$.
For simplicity, we will consider the  physical case of a wave generated in a region of zero current, $U(0)=0$, that enters into 
a region where an opposite current starts increasing its speed (in absolute value) and then adjusts to 
some constant value $U_0$. Note that in this case the coefficient of the nonlinear term of 
equation ($\ref{NLSM}$) increases as the waves enter into the current  up to a certain value and then remains constant. The net effect is therefore an increase of the nonlinearity of the system. Numerical simulations of the current modified NLS equation presented in  \cite{onorato2011} showed that an envelope of an initially stable wave train becomes unstable  after entering in the current region. As a result, the maximum amplitude shows a growing trend for increasing the ratio $U_0/c_g$, corroborating the idea that an originally stable plane wave is transformed into a breather by the presence of a current.

In \cite{ruban2012nonlinear} it was noted that the modified NLS equation proposed in \citep{hjelmervik09} and used in \cite{onorato2011} does not preserve wave action  (see  \cite{gerber87}). To the lowest order, conservation of wave action can be accounted for in equation (\ref{NLSM}) by simply multiplying the ratio $U_0/c_g$ by a factor 2. Based on this modification, a prediction for the maximum wave amplitude obeys the following equation:
\begin{equation}
\frac {A_{max}} {\sqrt{E}} = 
1 + 2 \sqrt{ 1-\bigg[\frac{\exp\left(U_0/c_g\right)} {\sqrt{2}\varepsilon N}\bigg]^2 },
\label{predictionmiguel}
\end{equation}    
where $A_{max}$ is the maximum wave amplitude achieved in the region of constant current and $\sqrt{E}$ is standard deviation of the wave envelope once the current has reached its maximum constant value. In \cite{ruban2012nonlinear} a derivation of a modified NLS equation based on 
an Hamiltonian formulation of surface gravity waves has been performed. 
 A similar prediction as 
the one in (\ref{predictionmiguel}) has been proposed and takes the following form:
\begin{equation}
\frac {A_{max}} {\sqrt{E}} = 
1 + 2 \sqrt{ 1-\bigg[\frac{(1+\sqrt{1+2 U_0/c_g})^{4} }{\sqrt{2}\varepsilon N 16 (1+2 U_0/c_g)^{1/4}} \bigg]^2 }.
\label{predictionruban}
\end{equation}  
Note that, to the lowest order, an expression equivalent to (\ref{predictionruban}) can be derived from equation (\ref{NLSM}) by multiplying the ratio $U_0/c_g$ by a factor 3. 

\begin{figure}[h]
\includegraphics[width=13cm]{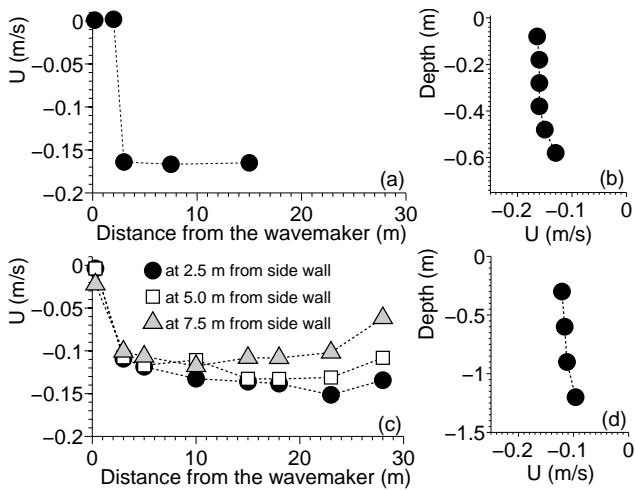}
\caption{Average horizontal velocity for a current field opposing wave propagation: longitudinal (a) and vertical (b) profiles in the wave flume of Plymouth University; and longitudinal (c) and vertical (d) profiles in the directional wave basin of the University of Tokyo. The longitudinal profiles were measured at a depth of 0.3 $m$ at the University of Tokyo and 0.08 $m$ at Plymouth University.}\label{tank}
\end{figure}

In this Letter we present two independent sets of laboratory experiments that were conducted in the wave flume of Plymouth University and in the narrow directional wave basin at the Ocean Engineering Tank of the University of Tokyo. 
The wave flume at Plymouth University is 35 $m$ long and 0.6 $m$ wide with a uniform water depth of 0.75 $m$. The facility is equipped with a piston wavemaker with active force absorption at one side and a passive absorber panel at the other end; only unidirectional propagation is allowed. The flume is also equipped with a pump for the generation of a background current up to 0.5 $m/s$, which can follow or oppose the wave direction of propagation (only an opposing current was used for the present study though). One of the inlet/outlet is located nearby the absorber, while the other is at a distance of about 2.5 $m$ from the wavemaker. This particular configuration allows waves to be generated outside the current field and propagate for a few wavelengths before encountering a current gradient. The wave field was monitored with 10 wave probes equally spaced along the tank, while the velocity field was monitored with two Acoustic Doppler Velocimeters (ADV) properly seeded. A survey of the current revealed a fairly uniform flow both in space and time. An example of longitudinal and vertical profiles of the average horizontal velocity is presented in Figs. \ref{tank}a and \ref{tank}b.

The Ocean Engineering Tank of the Institute of Industrial Science, University of Tokyo (Kinoshita Laboratory and Rheem Laboratory), is 10 $m$ wide, 50 $m$ long and 5 $m$ deep. It is equipped with a multidirectional wavemaker with 32 triangular plungers (0.31 $m$ wide) \cite{waseda09}. A sloping beach is deployed opposite the wavemaker to absorb the wave energy. The tank is also equipped with a pump (located beside the basin) for the generation of background currents up to 0.4 $m/s$; the stream can follow or oppose the waves. One of the inlet/outlet is located below the beach, while a second is located below the wavemaker (approximately 2 $m$ below the water level). Note that no modification of the cross section was performed to locally modify the velocity field. Wave probes were deployed along the tank at a distance of 2.5 $m$ from the sidewall and arranged at 5 $m$ intervals to monitor the evolution of wave trains (a six-probe array was also deployed to monitor directional properties); for consistency with the wave flume experiments, only probes within 25 $m$ from the wavemaker were considered, though. Two electromagnetic velocimeters were used to survey the current. Instruments were deployed at several locations in the tank at a depth of 0.2 $m$; a vertical profile was measured at about 10 $m$ from the wavemaker. Instantaneous measurements of horizontal velocity revealed that the flow had substantial spatial and temporal variations, with a dominant oscillation period of approximately 150 $s$. Average values are presented in Figs. \ref{tank}c and \ref{tank}d. For these tests,  the standard deviation was about 25\% of the mean over the entire time series. As the flow's outlet is located just below the wave generator, the velocity is approximately zero at a distance of about 0.2 $m$ from the wavemaker, while the flow is at regime at a distance of 5 $m$ from the wavemaker. Waves are therefore generated in a condition of (almost) still water and undergo a current gradient about 1 $m$ after being generated. Farther from the wavemaker, between 5 and 30 $m$ from the generator, the current still shows a weak gradient, which may slightly affect the wave field. Although the average horizontal velocity weakly decreased with the water depth, the vertical profile remained fairly uniform over a depth of about 1.5 $m$. The survey of the current field also indicated that the current runs faster on one side (along the wave probes), while it is slower on the other (see Fig. \ref{tank}c, for example).
It is important to mention that the cross-tank gradient refracts the wave field towards the sidewall (probes side). As such, this may cause linear directional focusing, steepening the wave profile and hence affecting wave dynamics.

A number of tests  characterized by different values of the modulation frequency and current velocity have been carried out in both facilities. For all tests, the initial signal at the wavemaker consisted of a three-component system: a carrier wave of period $T_0 = 0.8$ $s$ (wavelength $\lambda_0=2\pi/k_0\simeq1$ $m$) and two side bands with amplitudes $b_{\pm}$ equal to 0.25 times the amplitude $a_c$ of the carrier waves. 
Considering the water depth of 0.75 $m$ in the wave flume and 5 $m$ in the directional basin, experiments were performed under deep water conditions ($k_0h >4$). As the effect of the current is to 
steepen the wave profile, a small initial steepness was selected in order to avoid wave breaking. The tests presented here  were carried out by selecting the wave amplitude of the carrier wave  $a_c$  in such a way that the wave steepness was $k_0 a_0 = 0.063$ with $a_0^2 = a_c^2 +b_+^2 + b_-^2$. The frequency of the disturbances was chosen to force the number of waves under the perturbation $N = \omega_0 / \Delta \Omega$ (with $\omega_0$ being the angular frequency of the carrier waves) to be equal to 11. Under these circumstances, the perturbation frequency lays just outside the NLS-based instability region, i.e. waves are stable ($\varepsilon N = k_0 a_0 N =0.69 < 1/\sqrt{2}$). 
These packets were tested against different opposing current speeds. Velocities ranged from 0 $m/s$ to -0.30 $m/s$ with step of -0.02 $m/s$. All tests were run for a time period of 10 minutes. Considering the variability of the current in the directional basin, this ensured enough data to perform a standard statistical analysis of the observations. 

\begin{figure}
\includegraphics[width=9cm,height=11cm]{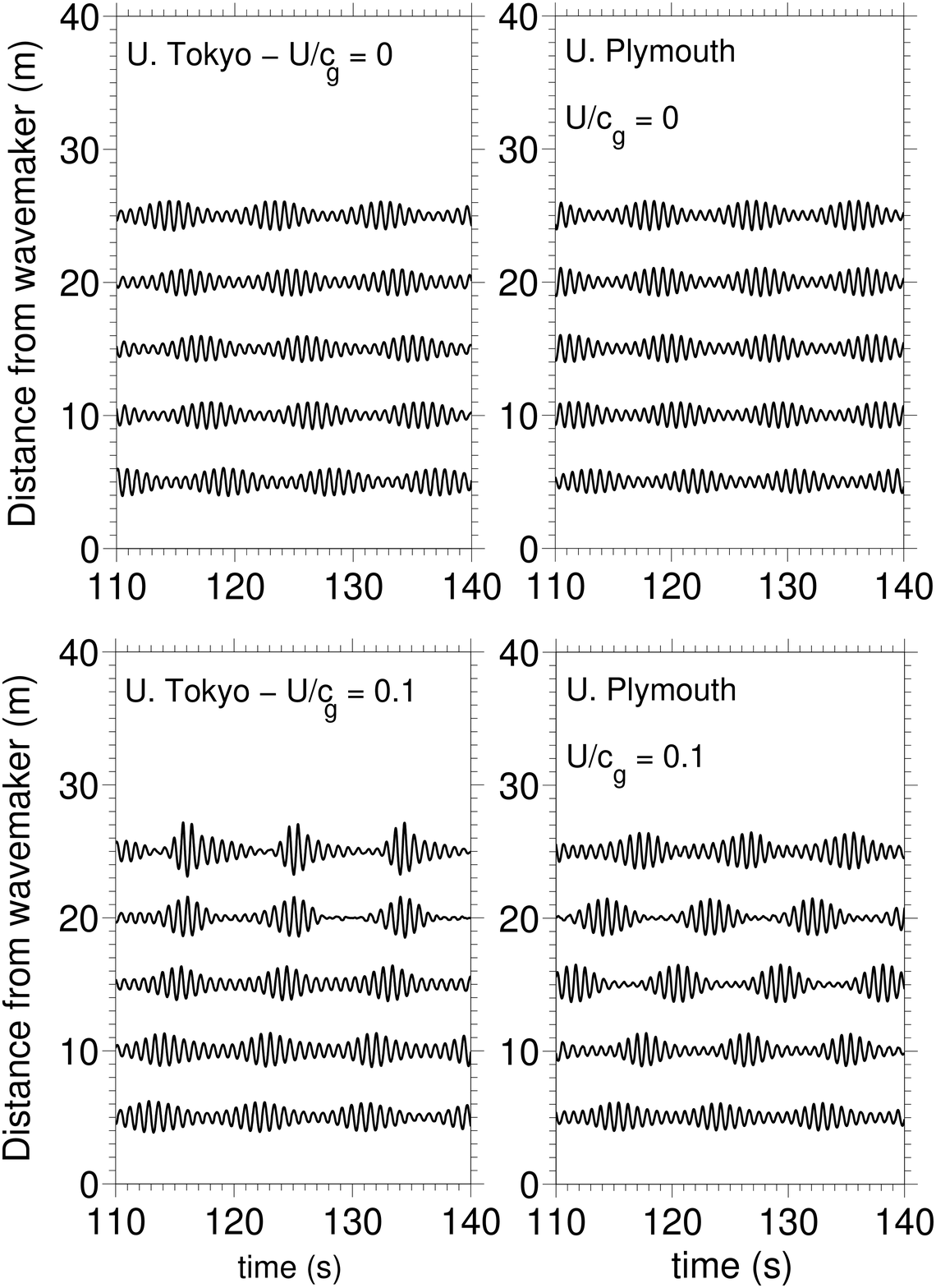}
\caption{Evolution of surface elevation: samples from the wave flume of Plymouth University (right panels) and the directional wave basin of the University of Tokyo (left panels).}\label{elev}
\end{figure}

Typical time series of the water surface elevation are presented in Fig. \ref{elev} for $U_0/c_g = 0$ and $U_0/c_g = -0.1$. What is clear from the figure is that, while the modulation does not grow in absence of current (top panels), the wave packets undergo a modulationally unstable process (nonlinear focussing) and wave amplification in the presence of a current. Interestingly enough, amplification is less sharp in the wave flume due to a more regular current field and the absence of contaminating three-dimensional effects.
%

\begin{figure}
\includegraphics[width=9cm]{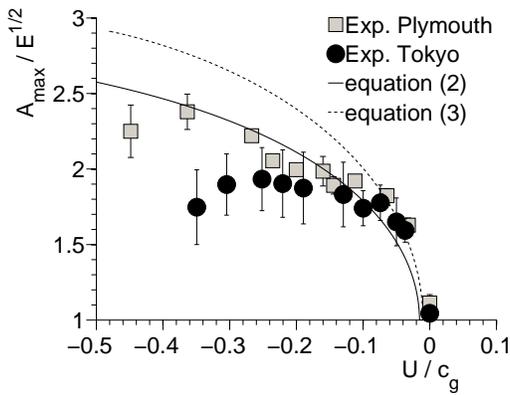}
\caption{Normalised maximum amplitude as a function of $U_0/c_g$: 
equation (\ref{predictionmiguel}), solid line; equation (\ref{predictionmiguel}) modified
by including the conservation of wave action, dashed line.}\label{amp}
\end{figure}

A standard zero-crossing procedure was applied to extract the maximum amplitude at each probe. Because of the temporal variability of the current in the directional basin, the analysis was performed on segments of three consecutive wave groups, i.e. time window $\tau = 26.4s$, where the current was nearly steady (velocity fluctuations were confined within a range of $\pm0.01$ $m/s$, namely one order of magnitude smaller than the average). As the prediction in equation (\ref{predictionmiguel}) only includes the contribution of free wave modes, frequencies greater than $1.5 \:\omega_c$ and smaller than $0.5 \:\omega_c$ were removed to filter out bound modes. The amplitude was then normalised by $E^{1/2} = [(1/\tau)\: \int_0^{\tau}{|A|^2 dt}]^{1/2}$, where $A$ is the wave envelope of the concurrent segment, to eliminate the current-induced increase of wave amplitude. An average normalised maximum amplitude and standard deviation were calculated over the entire time series. The maximum wave amplitude is presented as a function of $U_0/c_g$ in Figure \ref{amp} together with equation (\ref{predictionmiguel}) (solid line) and the prediction model in (\ref{predictionruban}) (dashed line); error bands equivalent to the 68\% confidence interval (one standard deviation) are also shown; owing to the stable current field, the error band for flume experiments is notably smaller than the one detected in the directional basin. Qualitatively, both tests are in good agreement with theory, substantiating that an adverse current gradient triggers moduational instability processes. Quantitatively, however, observations are notably overestimated by the model in (\ref{predictionruban}), especially for strong current fields. Equation (\ref{predictionmiguel}), on the contrary, produces a satisfactory approximation of the records, particularly for the flume experiments. It is important to remark, in this regard, that three dimensional effects induced by refraction enhance the breaking probability in the directional wave basin and hence limit the maximum wave amplitude also for relatively mild currents. Amplitude growth ceases, in fact, for $| U_0 / c_{g} |  > 0.2$ as individual waves systematically reach their limiting steepness \cite{toffoli10} and break. In the wave flume, three-dimensional effects are suppressed, making breaking dissipation less likely and thus allowing waves to develop the maximum amplification. This justifies the good agreement with theory until $| U_0 / c_{g} |  \approx 0.4$; beyond this threshold waves reach the limiting steepness and break. 


In conclusion, the conjecture on the generation of rogue waves in opposite current using a current-modified cubic NLS equation has been here confirmed by two independent sets of laboratory experiments, which were carried out in a wave flume and a narrow directional wave basin. Observations corroborate that the excitation of the modulation and the concurrent intensification of the maximum wave growth is a function of the adverse current  $U_0 / c_{g}$ and is consistent with theoretical predictions in equation (\ref{predictionmiguel}), but  overestimated by the model in (\ref{predictionruban}). 
Despite possible shortcomings of the theory and uncertainties in the experimental conditions, especially related to the irregularities of the current field, the present study clearly shows evidence that
an opposing current field can destabilize an otherwise stable wave packet. The result is the formation of a rogue wave, whose amplitude depends on the ratio of current velocity to group velocity.  Note that the essence of the theoretical predictions in \cite{onorato2011} and \cite{ruban2012nonlinear} is to account for the impact of the current shear only to modify the basic parameters that control the dynamics of the wave train. Once the basic parameters have changed, the instability process follows its nominal evolution pattern (namely, the nonlinear stages of modulational instability). It is important to mention that  this is the case for our experiments because the spatial scale of the current field variation was smaller than the spatial scale of the nonlinear evolution. This is not necessarily the case in reality and further research is needed for those cases when the two scales are of the same order. 
From a physical point of view, the process that we have described may take place in nature when a modulationally stable swell (waves propagating without the forcing of the wind), which is characterized by a narrow spectrum (both in direction and frequency) enters a region  of an opposite variable current. Besides the linear effect of refraction which could generate  linear focussing, the current gradient can destabilise the wave packets leading to  a nonlinear focussing effect and the formation of  rogue waves.  Indeed, as mentioned in \cite{onorato2011}, such currents may reach velocities up to 1.5 $m/s$, and for a group velocity corresponding to waves of period equal to 10 $s$ (a typical condition during storms), the ratio $U_0 /  c_g$ is of the order of 0.2, large enough to trigger a dangerous rogue wave.  We believe that the mechanism we have observed is universal and  can be reproduced in all  fields  of physics where an NLS equation with variable coefficient of the nonlinear term can be written.

{\bf Acknowledgments} Experiments were supported by the JSPS Fellowship for Research in Japan Program, Grants-in-Aid for Scientific Research of the JSPS and the International Science Linkages (ISL) Program of the Australian Academy of Science. Financial support of the Australian Research Council and Woodside Energy Ltd through the grant LP0883888 is also acknowledged. M. O. and A. T. acknowledge the VDRS grant from Swinburne University of Technology.  M.O. was supported by EU project EXTREME SEAS (SCP8-GA-2009-234175)
and ONR grant N000141010991. M.O. acknowledges Dr. B. GiuliNico  for interesting discussions.
\bibliography{references}
\end{document}